\documentclass [prl,aps,letterpaper,twocolumn,superscriptaddress,amsmath,amssymb,floatfix] {revtex4}

\usepackage{graphicx}
\usepackage{textcomp}
\usepackage{epstopdf}
\usepackage{mathptmx}

\usepackage[latin9]{inputenc}
\usepackage{amsmath}
\usepackage{amssymb}
\usepackage{graphicx}
\usepackage{esint}

\usepackage[unicode=true,pdfusetitle,
 bookmarks=true,bookmarksnumbered=false,bookmarksopen=false,
 breaklinks=false,pdfborder={0 0 0},pdfborderstyle={},backref=false,colorlinks=true]
 {hyperref}
\hypersetup{
 colorlinks,linkcolor=red,citecolor=blue}

\makeatletter

\usepackage{xcolor}\usepackage{soul}

\newcommand{\ket}[1]{\ensuremath{\left|#1\right\rangle}}
\definecolor{blue}{rgb}{0,0,1}
\definecolor{red}{rgb}{1,0,0}
\definecolor{green}{rgb}{0,1,0}

\makeatother

\begin{document}

\title{Heisenberg limited single-mode quantum metrology}

%
%
%
%
%

\author{W.~Wang}
\thanks{These two authors contributed equally to this work.}

\affiliation{Center for Quantum Information, Institute for Interdisciplinary
Information Sciences, Tsinghua University, Beijing 100084, China}

\author{Y.~Wu}
\thanks{These two authors contributed equally to this work.}

\affiliation{Center for Quantum Information, Institute for Interdisciplinary
Information Sciences, Tsinghua University, Beijing 100084, China}
\affiliation{Department of Physics, University of Michigan, Ann Arbor,
Michigan 48109, USA}

\author{Y.~Ma}
\affiliation{Center for Quantum Information, Institute for Interdisciplinary
Information Sciences, Tsinghua University, Beijing 100084, China}

\author{W.~Cai}
\affiliation{Center for Quantum Information, Institute for Interdisciplinary
Information Sciences, Tsinghua University, Beijing 100084, China}

\author{L.~Hu}
\affiliation{Center for Quantum Information, Institute for Interdisciplinary
Information Sciences, Tsinghua University, Beijing 100084, China}

\author{X.~Mu}
\affiliation{Center for Quantum Information, Institute for Interdisciplinary
Information Sciences, Tsinghua University, Beijing 100084, China}

\author{Y.~Xu}
\affiliation{Center for Quantum Information, Institute for Interdisciplinary
Information Sciences, Tsinghua University, Beijing 100084, China}

\author{Zi-Jie Chen}
\affiliation{Key Laboratory of Quantum Information, CAS, University of
Science and Technology of China, Hefei, Anhui 230026, P. R. China}

\author{H. Wang}
\affiliation{Center for Quantum Information, Institute for Interdisciplinary
Information Sciences, Tsinghua University, Beijing 100084, China}

\author{Y.~P.~Song}
\affiliation{Center for Quantum Information, Institute for Interdisciplinary
Information Sciences, Tsinghua University, Beijing 100084, China}

\author{H.~Yuan}
\affiliation{Chinese University of Hong Kong, Hong Kong, China}

\author{C.-L.~Zou}
\affiliation{Key Laboratory of Quantum Information, CAS, University of
Science and Technology of China, Hefei, Anhui 230026, P. R. China}

\author{L.-M. Duan}
\affiliation{Center for Quantum Information, Institute for Interdisciplinary
Information Sciences, Tsinghua University, Beijing 100084, China}

\author{L.~Sun}
\affiliation{Center for Quantum Information, Institute for Interdisciplinary
Information Sciences, Tsinghua University, Beijing 100084, China}



\begin{abstract}
Two-mode interferometers, such as Michelson interferometer
based on two spatial optical modes, lay the foundations for quantum
metrology~\cite{Giovannetti2011,Pezze2018,Braun2018}. Instead of
exploring quantum entanglement in the two-mode interferometers, a
single bosonic mode also promises a measurement precision beyond the
shot-noise limit (SNL) by taking advantage of the infinite-dimensional
Hilbert space of Fock states~\cite{Jacobs2017}. However, the experimental
demonstration still remains elusive. Here, we demonstrate a single-mode phase estimation that approaches the Heisenberg limit (HL) unconditionally. Due to the strong dispersive nonlinearity and long coherence time of
a microwave cavity, quantum states of the form $\left(\left|0\right\rangle +\left|N\right\rangle \right)/\sqrt{2}$
are generated, manipulated and detected with high fidelities, leading
to an experimental phase estimation precision scaling as $\sim N^{-0.94}$. A $9.1$~$\mathrm{dB}$
enhancement of the precision over the SNL at $N=12$, which is only
$1.7$~$\mathrm{dB}$ away from the HL, is achieved. Our experimental
architecture is hardware efficient and can be combined with the quantum
error correction techniques to fight against decoherence~\cite{Ofek2016,Hu2018},
thus promises the quantum enhanced sensing in practical applications.
\end{abstract}

\maketitle

\begin{figure}
\includegraphics{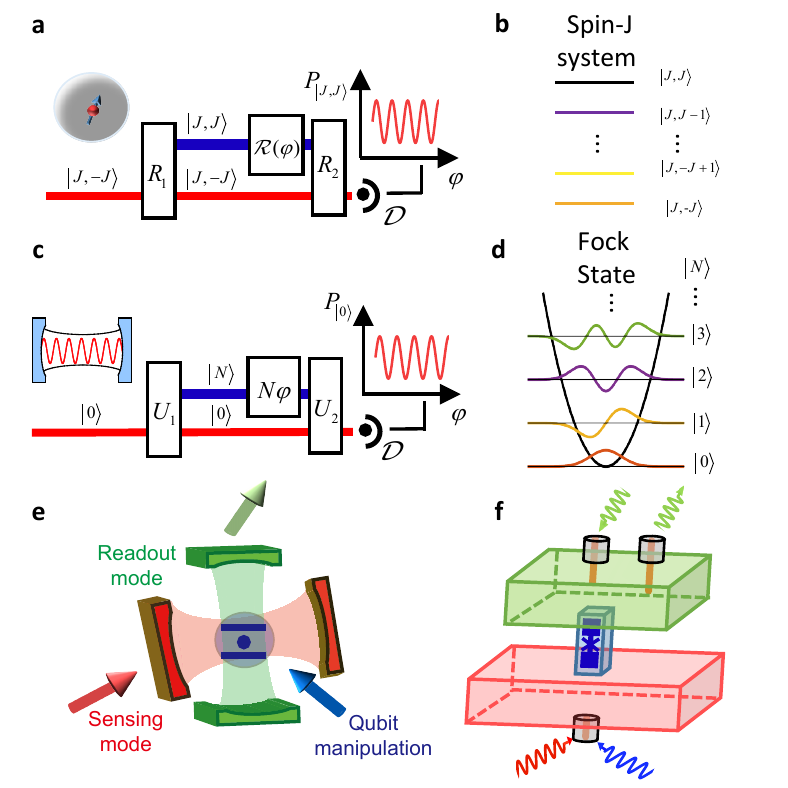} \caption{\textbf{Single-mode quantum metrology architectures.} \textbf{a} and
\textbf{b} Single-atom Ramsey interferometer with a total angular
momentum number $J$. The best precision can be achieved by using
the superposition of $\left|J,-J\right\rangle $ and $\left|J,J\right\rangle $
with the maximum variance of angular momentum. \textbf{c} and \textbf{d}
Single-mode Ramsey interferometer for photons/bosons with an optimal
precision achieved by the superposition of Fock states with the maximal
variance of photon numbers for a given mean photon number. \textbf{e}
and \textbf{f} Schematic illustrations of the single-mode sensing
architecture and the experimental circuit quantum electrodynamics
system. A single qubit couples to two photonic cavity modes, with
the two modes serving as the sensing mode and the ancillary qubit
readout mode, respectively. The two boxes represent the microwave cavities,
between which the ancillary superconducting qubit on a chip is located
in a waveguide trench.}
\label{fig:figure1} \vspace{-6pt}
\end{figure}

Based on the coherent interference effects, interferometers have been
extensively used in precision measurements. For example, the two-mode
atomic Ramsey interferometer that manipulates the superpositions of
two internal states of an atomic ensemble has been used in various
applications, such as clock, gravimeter, and gyro~\cite{Chu2002}.
Similarly, by separating photons into two spatial modes, two-mode
photonic Michelson interferometers have been extensively used in LIGO~\cite{Adhikari2014},
optical coherence tomography and spectrometer. Recently, quantum metrology,
which makes use of quantum mechanical effects, such as entanglement,
has gained a lot of attention in the two-mode interferometers, as
it can achieve measurement precisions beyond the classical limit~\cite{Giovannetti2004sci,Schnabel2010,Giovannetti2011,Pezze2018,Braun2018}.
In the applications of quantum metrology, highly entangled states,
such as the Greenberger\textendash Horne\textendash Zeilinger state
of an atomic ensemble~\cite{Pezze2018,Braun2018} or the NOON state of optical interferometer~\cite{Nagata2007Science,Slussarenko2017},
are essential. To prepare these exotic quantum states, non-local operations
are required. In addition, the optimal measurements are also typically
highly non-local. This poses significant challenges for practical applications
of quantum metrology.


\begin{figure*}
\includegraphics{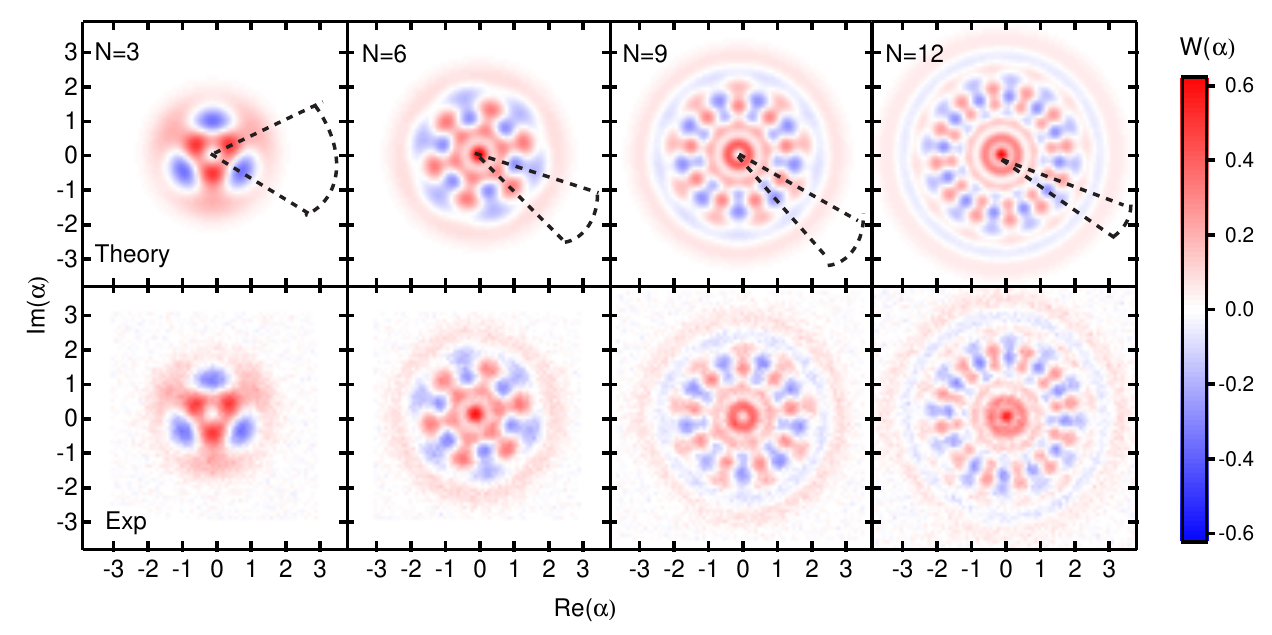} \caption{\textbf{Wigner functions of the maximum variance states.} Theoretical
(top panel) and experimental (bottom panel) results for $N=3,6,9,12$
with state fidelities of 0.94, 0.92, 0.83, 0.70, respectively.
The measured state preparation fidelity decays with $N$, mainly attributed
to the larger probability of photon loss and the worse reconstruction
measurement to obtain the Wigner functions for larger $N$. The measured range of the real and imaginary
parts of $\alpha$ is {[}-3.0,2.9{]} for $N=3$ and 6, {[}-3.6,3.5{]}
for $N=9$, and {[}-3.9,3.8{]} for $N=12$. The angles of the dashed
sectors indicate that the sensitivity of phase estimation scales as
$1/N$.}
\label{fig:wigner} \vspace{-6pt}
\end{figure*}



In this Letter, instead of exploring quantum entanglement in the two-mode interferometer we implement the single-mode photonic quantum metrology with a superconducting qubit-oscillator system~\cite{DevoretSchoelkopf} and demonstrate an unconditional phase estimation with the precision approaching the HL. A quantum sensor with a single mode is of great
interest~\cite{Duivenvoorden2017,Jacobs2017} for its hardware efficiency, compactness, and robustness against non-local perturbations. For a single mode, the phase can be measured based on the photon number
dependent phase accumulation. Using the state $\ket{\Psi\left(N\right)}=\left(\left|0\right\rangle +\left|N\right\rangle \right)/\sqrt{2}$, superpositions of Fock states, up to $N=12$, we demonstrate a phase estimation precision which scales as $\delta\tilde{\theta}\sim N^{-0.94}$ and approaches the HL. At $N=12$, $\delta\tilde{\theta}$ corresponds to an enhancement of $20\mathrm{log}_{10}({\delta\tilde{\theta}_{\mathrm{SNL}}}/{\delta\tilde{\theta}})~\mathrm{dB}= 9.1~\mathrm{dB}$ over the SNL $\delta\tilde{\theta}_{\mathrm{SNL}}$. Envisioning future applications in the optical regime with microwave-to-optical transduction, we also realize a measurement scheme that is easy to implement in optics and only uses displacement operations and photon counting. Under
this restricted measurement scheme, an SNL-beating precision, which
scales as $\delta\tilde{\theta}\sim N^{-0.62}$, is also achieved.

According to the quantum Cramer-Rao bound~\cite{Braunstein1994},
the estimation precision of parameter $\theta$ encoded in the state
$|\psi(\theta)\rangle=e^{-i\theta H}\left|\psi\right\rangle $ is
lower bounded as $\delta\tilde{\theta}\geq\frac{1}{2\Delta H}$,
where $\delta\tilde{\theta}$ is the standard deviation of an unbiased
estimator $\tilde{\theta}$, and $(\Delta H)^2=\left\langle \psi\right|H^{2}\left|\psi\right\rangle -\left\langle \psi\right|H\left|\psi\right\rangle ^{2}$
is the variance of the Hamiltonian $H$ under the initial probe $|\psi\rangle$.
The quantum states with a maximum variance therefore are optimal for
the single-mode sensing, i.e. the equal superpositions of the eigenstates
of $H$ corresponding to the extreme eigenvalues are the most preferable
quantum states. For example, as illustrated in Figs.$\,$\ref{fig:figure1}\textbf{a}
and \ref{fig:figure1}\textbf{b}, an atom prepared in the equal
superposition of angular momentum states $\left|J,-J\right\rangle $
and $\left|J,J\right\rangle $ has maximal sensitivity to external
field ($H=J_{z}$ and $J_{z}$ is the angular momentum operator).
Recently, a high precision electrometer, using the Schr\"{o}inger cat
state of large angular momentum states to enhance $\Delta H$, has
also been demonstrated to beat SNL~\cite{Facon2016,Chalopin2018,Dietsche2019}.
Similarly, the phase precision with a single bosonic mode would be
enhanced by using the state $\ket{\Psi\left(N\right)}=\left(\left|0\right\rangle +\left|N\right\rangle \right)/\sqrt{2}$,
since it has the maximum variance for $H=a^{\dagger}a$ ($a$ is the
bosonic operator of the sensing mode) (Figs.$\,$\ref{fig:figure1}\textbf{c}
and \ref{fig:figure1}\textbf{d}). Such a maximum variance state (MVS)
can in principle achieve the HL precision $\delta\tilde{\theta}=1/N$
with $\sqrt{N}$ times enhancement over the SNL.

As schematically illustrated in Figs.$\,$\ref{fig:figure1}\textbf{e}
and \ref{fig:figure1}\textbf{f}, our experiment is carried out with
a superconducting system consisting of a transmon qubit dispersively
coupled to two three-dimensional cavities~\cite{Paik,Ofek2016,Hu2018}.
One long-lived cavity serves as the sensing mode; the transmon qubit
as an ancilla assists the preparation, manipulation and detection
of the photonic states in the sensing mode; the short-lived cavity
is employed for a high-fidelity readout of the qubit state. The Hamiltonian
of the qubit-oscillator system is $H=-\hbar\chi_{\mathrm{qs}}a^{\dagger}a|e\rangle\langle e|$~\cite{DevoretSchoelkopf},
where $|e\rangle$ is the excited state of the qubit (the ground state
is $\ket{g}$), and $\chi_{\mathrm{qs}}$ reflects the dispersive
interaction strength between the qubit and the mode. In our system,
$\chi_{\mathrm{qs}}/2\pi=1.90\,\mathrm{MHz}$ is much stronger
than the decoherence rates of the qubit and the sensing
mode, thus allows full control of the photonic quantum state~\cite{Heeres2015,Ofek2016,Heeres2017,Wang2017,Hu2018}.


\begin{figure*}
\includegraphics{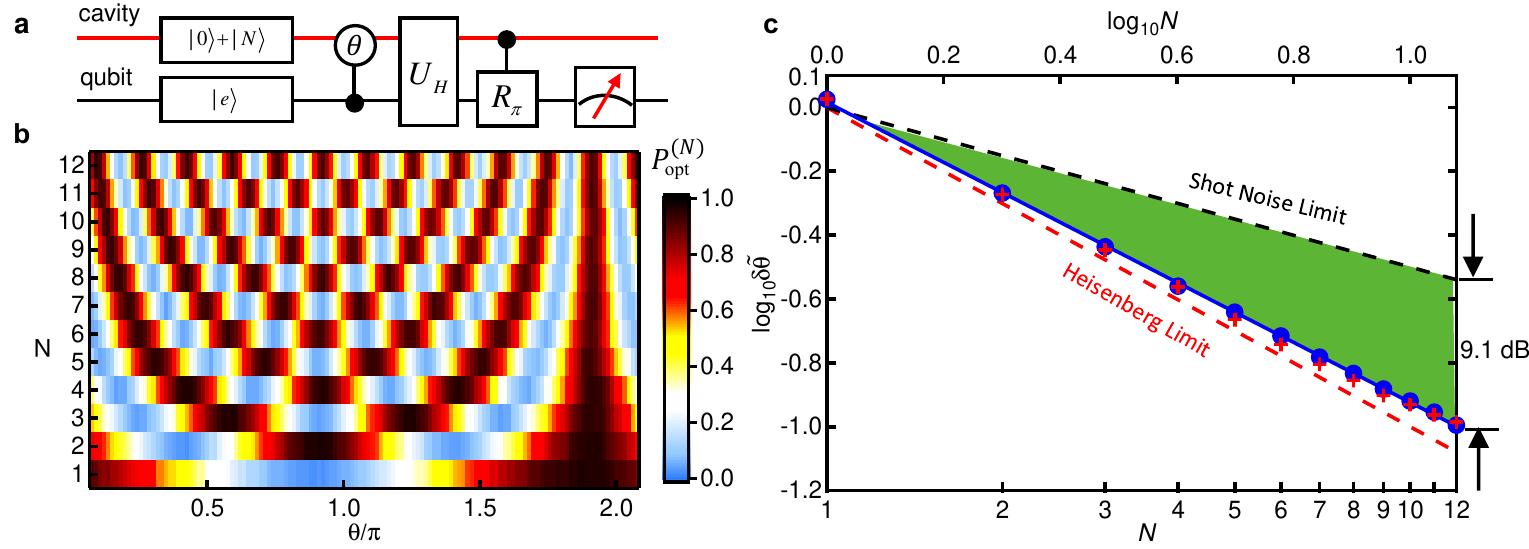} \caption{\textbf{Optimal single-mode sensing scheme.} \textbf{a} Quantum circuit.
\textbf{b} The measured probability $P_{\mathrm{opt}}^{(N)}$ of projecting
to $\left|\Psi\left(N\right)\right\rangle $ as a function of the phase
$\theta$ in the optimal scheme for different $N$. The outcomes are obtained with $10^{6}$ repetitions of the experiment. \textbf{c} Quantum advantage of the optimal scheme. Blue dots are experimental results.
The blue solid line is a linear fit and gives $\log_{10}\delta\tilde{\theta}=-0.94\log_{10} N+0.016$
with the precision scaling $N^{-0.94}$ approaching the Heisenberg
scaling ($N^{-1}$). The small offset $0.016$ in the log-log scale of the precision
is dominantly due to the imperfections of the qubit readout process,
where the spontaneous decay of the qubit gives wrong indication
of the cavity state and lowers the contrast of the Ramsey interference.
The red crosses are the results from numerical simulations including
the decoherences of our system, in good agreement with the measured
data. The black and red dashed lines are the theoretically calculated
SNL and HL, respectively. Green region represents the experimental results that
surpass the standard limit by about 9.1~dB at $N=12$. Error bars
are smaller than the markers.}
\label{fig:optimal_scheme} \vspace{-6pt}
\end{figure*}

In our experiment, the probe states of the sensing mode are deterministically
created by implementing a qubit-assisted unitary operation on the
mode. With numerically optimized control pulses~\cite{Khaneja2005},
the probe states $\ket{\Psi\left(N\right)}=\left(\left|0\right\rangle +\left|N\right\rangle \right)/\sqrt{2}$ with $N=1,2,\dots,12$
are prepared faithfully. In Fig.~\ref{fig:wigner} the experimentally
measured Wigner functions (bottom panels) of the typical MVSs are
plotted, agreeing well with the ideal ones (top panels). 
In the phase-space, there are interesting periodic fringes in the
polar direction with $N$-fold rotational symmetry for $\ket{\Psi\left(N\right)}$.
As the rotation of the Wigner function by $\theta$ corresponds to
the phase operation $U\left(\theta\right)=e^{i\theta a^{\dagger}a}$
on the oscillator, the enhanced measurement precision by the MVS can
be intuitively explained: because of the fine fringe features, $\left|\Psi\left(N\right)\right\rangle$ would be rotated to an orthogonal
state when the phase $\theta=\pi/N$, the measurement precision with
the MVS is thus proportional to $N$, instead of $\sqrt{N}$. 

Figure~\ref{fig:optimal_scheme}\textbf{a} depicts the experimental
circuit for the optimal sensing scheme by a Ramsey-like interference
(Fig.$\,$\ref{fig:figure1}\textbf{c}), which is potential for attaining
the ultimate precision HL for the single-mode sensing. After an initialization
process, the cavity is prepared in $\left|\Psi\left(N\right)\right\rangle $
while the qubit ends up with $\ket{e}$, resulting in a phase operation
$U\left(\theta\right)$ on the sensing mode with $\theta=-\chi_{\mathrm{qs}}\tau$
and $\tau$ being the waiting time. Then, a unitary $U_{H}$ is implemented
to rotate $\ket{\Phi_{+}}=\ket{e}(\ket{0}+\ket{N})/\sqrt{2}$ to $\ket{g}\ket{0}$
and $\ket{\Phi_{-}}=\ket{e}(\ket{0}-\ket{N})/\sqrt{2}$ to $\ket{e}\ket{0}$.
Finally, the ancillary qubit is projectively measured on $\left|g\right\rangle $,
giving projection of $U\left(\theta\right)\left|\Psi\left(N\right)\right\rangle $
onto $\left|\Psi\left(N\right)\right\rangle $ with the ideal probability
oscillation $P_{\mathrm{opt}}^{(N)}=\left(1+\mathrm{cos}N\theta\right)/2$.


\begin{figure*}[t]
\includegraphics{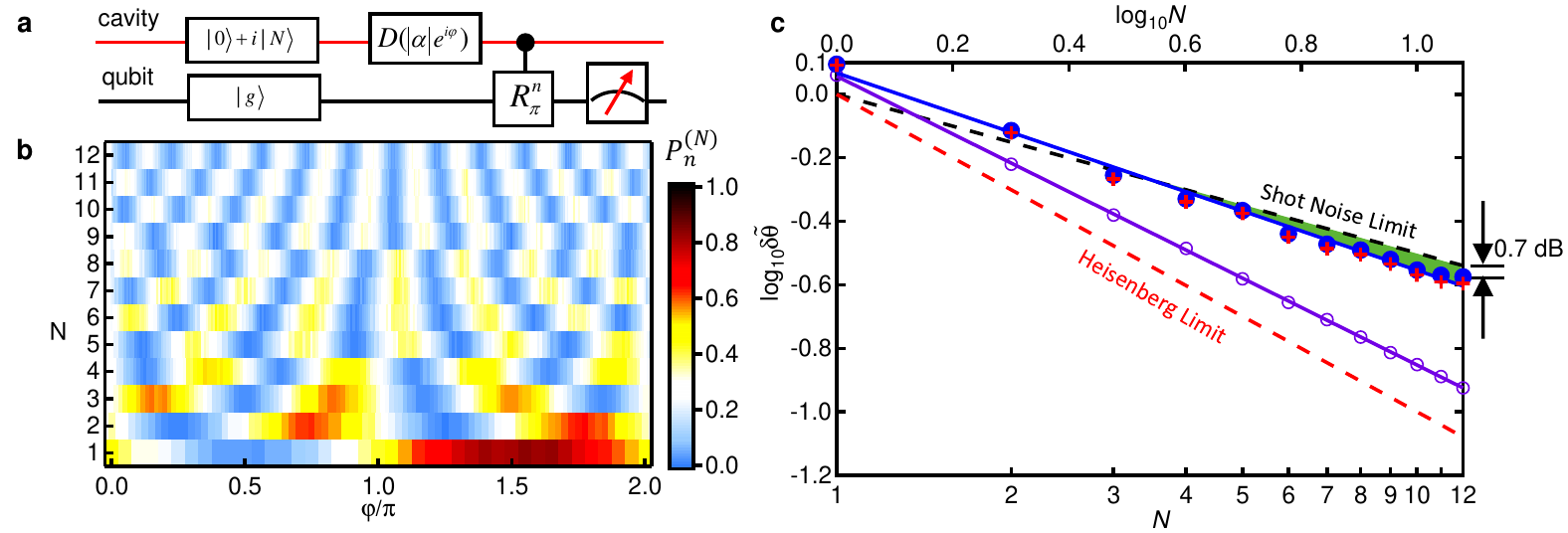} \caption{\textbf{Hybrid single-mode sensing scheme.} \textbf{a} Quantum circuit.
The phase operation $U\left(\varphi\right)=\mathrm{exp}(-i\varphi a^{\dagger}a)$
is effectively added to the sensing mode in the rotating-frame of
the displacement operation. The detection is realized by a combination
of a displacement of the photonic mode {[}$D(\alpha=|\alpha|e^{i\varphi})${]}
and selective photon number ($|n\rangle$ state) detection enabled
by the ancillary qubit through its dispersive interaction to photons.
\textbf{b} The measured probability $P_{n}^{(N)}$ of projecting to
the photon number state $\ket{n_{\mathrm{opt}}}$
as a function of $\varphi$ in the displacement operation, with the
optimal $n_{\mathrm{opt}}$ for each $N$ being numerically obtained.
\textbf{c} Quantum advantage for the hybrid scheme. Blue dots are
experimental results. The blue solid line is a linear fit and gives
$\log_{10}\delta\tilde{\theta}=-0.62\log_{10} N+0.068$. The red crosses are the
results from numerical simulations including the decoherences of our
system, in good agreement with the measured data. Green region represents
the experimental results that surpass the standard limit by about 0.7~dB
at $N=12$. Purple circles show the achievable precision for the hybrid
sensing scheme with a fit $\log_{10}\delta\tilde{\theta}=-0.91\log_{10} N+0.057$, provided all photon numbers can be detected. Error bars are smaller than the markers. The scaling $N^{-0.62}$ ($N^{-0.69}$
for an ideal experiment) can still beat the standard scaling $N^{-0.5}$
due to the initial MVS. The offset of the hybrid scheme is mainly due to the imperfect photon number detection.}
\label{fig:hybrid_scheme} \vspace{-6pt}
\end{figure*}

The experimental results of the optimal scheme $P_{\mathrm{opt}}^{\left(N\right)}$
are shown in Fig.$\,$\ref{fig:optimal_scheme}\textbf{b}. As intuitively
expected from Fig.$\,$\ref{fig:wigner}, the period of the Ramsey
interference fringes reduces with $N$ and the contrast of the fringes
are nearly ideal. By fitting the experimentally measured probability
with $P^{(N)}\left(\theta\right)=A+B\cos\left(N\theta\right)$, where
$A-B$ and $B$ represent the detected background and the contrast
of the Ramsey interference fringes, respectively, the phase estimation precision can be inferred as $\delta\tilde{\theta}=\sqrt{P^{(N)}\left(\theta\right)\left(1-P^{(N)}\left(\theta\right)\right)}/\frac{\partial P^{(N)}\left(\theta\right)}{\partial\theta}=\frac{\sqrt{A\left(1-A\right)}}{NB}$.

Figure~\ref{fig:optimal_scheme}\textbf{c} shows the results of $\delta{\tilde{\theta}}$
(blue dots) as a function of $N$ in a logarithmic-logarithmic scale.
Clearly, the optimal scheme beats the SNL, with the green region
representing the experimental results that surpass the SNL with a
maximum precision enhancement of 9.1~dB at $N=12$, which is only
$1.7$~$\mathrm{dB}$ away from the ultimate HL. The results demonstrate
the quantum advantage of our single-mode sensing unambiguously. The
obtained precision scales as $N^{-0.94}$, approaching the Heisenberg
scaling ($N^{-1}$). The slight deviation mainly attributes to the
$N$-dependent imperfections including the larger operation errors
for larger Hilbert space of Fock states (errors in the control pulse
and parameter uncertainties) and higher probability of photon loss.


The demonstrated optimal schemes can be utilized in practical sensing
applications, for example, the detection of the frequency and
power of a microwave signal based on the Stark-effect-induced phase shift of the sensing
mode. By utilizing the recently developed high-efficient bidirectional microwave-to-optical quantum transduction~\cite{Fan2018,Higginbotham2018}, our scheme with the MVS can also be employed for the optical metrology.
However, the Ramsey-like measurement is very challenging in optical
domain due to the limited capability of deterministic quantum state
manipulation of optical photons. We thus propose a hybrid sensing
scheme, as shown in Fig.$\,$\ref{fig:hybrid_scheme}\textbf{a}, by
employing a measurement scheme that only uses easy operations in
the optical domain, such as displacement operation and photon
counting~\cite{Matthews2016}. 

Envisioning the application of such a hybrid scheme, we simulate the
scheme in our superconducting system with the restricted measurement.
It is worth noting that different from photon counting in the
real optical system~\cite{Matthews2016}, the measurement through
the ancillary qubit in the superconducting system can only obtain
a binary output, i.e. a result of whether the photon number is $n$
or not. So, the outcome of the measurement has the probability $P_{n}^{(N)}=\left|\left\langle n\right|D\left(\alpha\right)U\left(\varphi\right)\left|\Psi\left(N\right)\right\rangle \right|^{2}$, where $D(\alpha)$ is the displacement operator and $U(\varphi)=\mathrm{exp}(-i\varphi a^{\dagger}a)$ is the phase operator.
We optimize the parameters $\alpha$ and $n$ for each MVS to maximize
the measurement precision (see Supplementary Materials).

The experimental results for the simulated hybrid scheme are summarized
in Fig.$\,$\ref{fig:hybrid_scheme}\textbf{b}. Although the fringe
period reduces with $N$ similar to that in the optimal scheme, the contrast
for the hybrid scheme reduces with $N$. The reason is mainly that
the probability of the binary photon number detection reduces for
large $N$ as the state spreads in the Fock space after a displacement
operation. However, this hybrid scheme beats the SNL as well, as indicated
by the green region in Fig.$\,$\ref{fig:hybrid_scheme}\textbf{c},
with a maximum precision enhancement of 0.7~dB at $N=12$. The obtained scaling
$N^{-0.62}$ ($N^{-0.69}$ for an ideal experiment) is lower than
that for the optimal scheme because of the sub-optimal detection process,
but can still beat the standard scaling $N^{-0.5}$ due to the initial
MVS. 
Actually, by using a photon number resolving detector, which is available
in optical domain, a better precision could be achieved by the hybrid
scheme in future optical sensing applications (as shown by purple
circles in Fig.~\ref{fig:hybrid_scheme}\textbf{c}).


Our single-mode quantum metrology architecture achieves a precision near the HL and holds the advantage of hardware efficiency, minimized sensing configuration, and compatibility with quantum error correction that can be employed for further enhancement of the precision~\cite{Zhou2018}. Our scheme can also be directly applied to other physical systems such as trapped ions~\cite{Zhang2018} and nitrogen-vacancy centers~\cite{Golter2016}. As demonstrated in the hybrid scheme, the precision still beats the SNL with the restricted detecting scheme consisting of only displacement operation and photon counting, which are easy to implement in optics. Additionally if we use microwave-to-optical up- and down-conversion twice, near-HL precisions with the optimal detecting scheme can be achieved. Our scheme thus also adds a powerful new platform to optical quantum metrology, which is quantum resource saving and robust compared to the multiple-path optical interferometer.


\begin{acknowledgments}
This work was supported by the National Key Research Funding No.2017YFA0304303
and the National Natural Science Foundation of China under Grant No.11474177.
H.Y. was supported by RGC Hong Kong(Grant No 14207717). C.-L.Z. was
supported by National Natural Science Foundation of China (Grant No.
11874342) and Anhui Initiative in Quantum Information Technologies
(AHY130000).
\end{acknowledgments}


%

\end{document}